\documentclass[trackchanges,twocolumn]{aastex701}
\usepackage{graphicx}
\usepackage{float}
\usepackage{amsmath}
\usepackage{amssymb}
\usepackage{subfigure}
\usepackage{cases}
\usepackage{mathrsfs}
\usepackage{threeparttablex}
\usepackage{longtable}
\usepackage{amssymb}
\usepackage{pifont}
\usepackage{epstopdf}
\usepackage{xcolor}
\usepackage[all]{hypcap}
\usepackage{mwe,tikz}
\usepackage{bm}
\usepackage{multirow}
\usepackage{xspace}
\usepackage{rotating}
\usepackage{CJK}
\usepackage{url}
\usepackage{upgreek}
\usepackage{booktabs}
\usepackage{textcomp}
\usepackage{makecell}
\usepackage{enumitem}
\usepackage{csquotes}
\usepackage{soul}
\usepackage{tabularx}
\usepackage{array}
\usepackage{xcolor}
\usepackage{bm}
\begin{document}

\title{Explosions in the Empty: A Survey of Transients in Local Void Galaxies}

\author[0009-0003-0474-7099]{Suo-Ning Wang}
\affiliation{School of Astronomy and Space Science, Nanjing University, Nanjing 210093, China, \\
\href{mailto:bbzhang@nju.edu.cn}{bbzhang@nju.edu.cn}}
\affiliation{Key Laboratory of Modern Astronomy and Astrophysics (Nanjing University), Ministry of Education, China}
\email{}

\author[0000-0003-4111-5958]{Bin-Bin Zhang}
\affiliation{School of Astronomy and Space Science, Nanjing University, Nanjing 210093, China, \\
\href{mailto:bbzhang@nju.edu.cn}{bbzhang@nju.edu.cn}}
\affiliation{Key Laboratory of Modern Astronomy and Astrophysics (Nanjing University), Ministry of Education, China}
\affiliation{Purple Mountain Observatory, Chinese Academy of Sciences, Nanjing, 210023, China}
\email{bbzhang@nju.edu.cn}

\author[0000-0002-7077-308X]{Rubén García-Benito}
\affiliation{Instituto de Astrof\'isica de Andaluc\'ia (CSIC), Glorieta de la Astronom\'ia s/n, 18008 Granada, Spain}
\email{}

\begin{abstract}
We present a systematic analysis of transient astrophysical events—including supernovae (SNe), gamma-ray bursts (GRBs), and fast radio bursts (FRBs)—in void and non-void galaxies within the local universe ($0.005 < z < 0.05$). Cosmic voids, defined by low galaxy densities and characterized by minimal environmental interactions, offer a natural laboratory for isolating the impact of large-scale underdensities on stellar evolution and transient production. Using multi-wavelength data from the Sloan Digital Sky Survey, the Sternberg Astronomical Institute Supernova Catalogue, and high-energy space observatories, we compare transient occurrence rates and host galaxy properties across environments. We find that core-collapse supernovae (CCSNe) are significantly more common in void galaxies, indicating that massive star formation remains active in underdense regions. In contrast, Type Ia supernovae are less frequent in voids, consistent with a scarcity of older stellar populations. Notably, we identify a short-duration GRB hosted by a void galaxy, demonstrating that compact object mergers can occur in isolated environments. Additionally, we find no FRBs associated with void galaxies. Taken together, these results show that cosmic voids exert a measurable influence on the star formation history of galaxies and hence on the production of transients.

\end{abstract}

\section{Introduction}
\label{section 1}

Cosmic voids are underdense regions occupying $\sim70$–$80\%$ of the cosmic volume while containing only $\sim10$–$15\%$ of the total matter (dark + baryonic) \citep{2012MNRAS.421..926P,2014MNRAS.442..462S,2014MNRAS.441.2923C,2018MNRAS.473.1195L,2019MNRAS.487.1607G}. Originating from primordial Gaussian perturbations, they evolve via gravitational instability as matter flows into surrounding overdensities; their reduced binding permits interior expansion exceeding the background Hubble flow \citep{2002PhR...367....1B,2014MNRAS.441.2923C,2024A&A...689A.213P}. Present-day individual voids span tens of Mpc in the local universe and form integral components of the cosmic web alongside clusters, filaments, and sheets \citep{2006Natur.440.1137S,2011IJMPS...1...41V,2016IAUS..308..493V,2018MNRAS.473.1195L}.

Void galaxies, although sparse, are statistically biased toward low masses, late morphologies, bluer colours, higher specific star formation rates (sSFR), and lower metallicities compared to galaxies in denser environments \citep{2005ApJ...624..571R,2006MNRAS.372.1710P,2007ApJ...658..898P,2012AJ....144...16K,2012MNRAS.425..641L,2014MNRAS.445.4045R,2016MNRAS.458..394B,2021ApJ...906...97F}. Their relative isolation—reduced ram pressure stripping and high-speed harassment \citep{1996Natur.379..613M,2022A&ARv..30....3B}—minimizes externally driven quenching, making voids natural laboratories for isolating internal drivers of star formation, chemical enrichment, and baryonic mass assembly \citep{2017NatAs...1E..40H}. 

Because transient progenitor channels are sensitive to stellar age, metallicity, and star formation intensity, the distinct demographic properties of void galaxies imply environmental modulation of specific transient classes--namely core-collapse supernovae (CCSNe), gamma-ray bursts (GRBs), and fast radio bursts (FRBs). Given their higher sSFRs and systematically lower metallicities, void galaxies should produce CCSNe and long GRBs (and potentially FRBs tied to recent star formation) at elevated rates, while yielding fewer Type Ia SNe per unit stellar mass. To quantify these environmental trends, we assemble matched samples of transients and host galaxies and compare their rates inside and outside voids. This quantitative assessment constrains how large-scale underdensity modulates transient production channels and informs models linking stellar population ecology to explosive endpoints.

The paper is organized as follows. Section~\ref{section 2} introduces the galaxy and transient samples and outlines our void--identification procedure. Section~\ref{section 3} details the environmental classification scheme and the computations of number and comoving--volume densities, as well as event rates. Section~\ref{section 4} discusses the implications and summarizes our conclusions. Throughout this work we adopt a cosmology with $H_{0}=73\,\mathrm{km\,s^{-1}\,Mpc^{-1}}$, $\Omega_{\mathrm{M}}=0.30$, and $\Omega_{\Lambda}=0.70$.

\section{Data and Methods}\label{section 2}

Our strategy is to fix a well‐defined cosmic volume, classify every galaxy within that volume as \emph{void} or \emph{non-void}, and then measure how the volumetric rates of several transient classes—CCSNe, Type Ia SNe, long/short GRBs, and FRBs—differ between the two environments. The procedure proceeds in four steps.

\begin{enumerate}[label=(\roman*)]
 \item \textbf{Define the void sample and redshift limit.} 
 We adopt the \textbf{\citet{2012MNRAS.421..926P}} SDSS VoidFinder catalog, which identifies 1\,055 voids with effective radii $\gtrsim10\,h^{-1}\,\mathrm{Mpc}$. To remain complete for both galaxies and transients, we restrict the analysis to the catalog’s reliable range \citep{2024A&A...689A.213P}, $0.005<z<0.05$. This slice therefore sets the redshift limits for \emph{all} subsequent samples.

 The redshift range of our sample was chosen as $0.005 \leq z \leq 0.05$. The lower limit ($z_{\min} = 0.005$) mitigates distance errors from peculiar velocities and prevents the shredding of extended nearby galaxies into multiple detections; stringent quality cuts (e.g., \texttt{class=`GALAXY'}) and visual inspections further ensure these are bona fide galaxies. The upper limit ($z_{\max} = 0.05$) ensures robust spectroscopic completeness, reliable derived physical parameters from the MPA--JHU catalog, and accurate host-galaxy associations for transients. We applied no additional magnitude cuts to avoid excluding faint galaxies that could host undetected transients.

\item \textbf{Galaxy census.} 
Using the SDSS DR8 spectroscopic database \citep{2011ApJS..193...29A} we extract every galaxy lying inside the cataloged right ascension/declination box and within $0.005 < z < 0.05$. Each galaxy is flagged as \emph{void} or \emph{non-void} according to its 3-D position relative to VoidFinder boundaries.

 \item \textbf{Transient compilation.} 
 For each transient class we query the relevant catalogs—\textcolor{black}{Sternberg Astronomical Institute Supernova Catalogue (SAI)} for SNe\citep{1993BICDS..42...17T}, the Greiner list for GRBs\footnote{\url{https://www.mpe.mpg.de/~jcg/grbgen.html}}, and the CHIME/FRB catalog (supplemented by \citealp{2025JCAP...01..036A}) for FRBs—and keep only events that 
 (a) fall inside the same right-ascension / declination limits, 
 (b) lie within $0.005 < z < 0.05$, and 
 (c) have secure host identifications. 
 Each host inherits the void/non-void label, yielding matched transient–galaxy pairs.

 \item \textbf{Rate and density calculations.} 
 From the matched sets we determine (a) number and comoving-volume densities, (b) volumetric event rates corrected for survey duration and sky coverage, and (c) subclass fractions (e.g., CCSNe vs.\ Type Ia, long vs.\ short GRBs). 
 These statistics quantify how large-scale underdensity influences the production of CCSNe, GRBs, and FRBs, thereby constraining models that connect stellar-population properties to energetic transient phenomena.

\end{enumerate}

\subsection{Galaxy and Void Catalogues}\label{subsec:galaxy_void}
%-------------------------------------------------
Our void sample is drawn from the SDSS VoidFinder catalogue of \citet{2012MNRAS.421..926P}, 
which identifies 79\,947 galaxies residing inside 1\,055 voids using the VoidFinder algorithm \citep{1997ApJ...491..421E,2002ApJ...566..641H}. 
The typical density contrast is $\delta\rho/\rho=-0.94\pm0.03$, and the minimum void radius is
$10\,h^{-1}\,{\rm Mpc}$.

Restricting the catalogue to our redshift interval $0.005<z<0.05$, and sky area $90^\circ<\mathrm{RA}<270^\circ$, $-4^\circ<\mathrm{Dec}<71^\circ$ yields\footnote{Our sample is set by the SDSS spectroscopic magnitude limit of $r<17.77$. Across our low-redshift range ($0.005<z<0.05$), the vast majority of void galaxies are sufficiently bright to be included, so overall completeness is not strongly affected. Residual selection effects—such as fiber collisions or reduced S/N near the flux limit—are expected to be small in underdense regions and are further mitigated by standard SDSS quality cuts (\texttt{zWarning=0}, \texttt{specPrimary=1}). Nonetheless, some intrinsically faint systems may be absent, which could lead to occasional host non-detections or misidentifications. We judge this effect to be minor and not to alter the qualitative conclusions of this work.} \[ N_{\rm gal}^{\rm void}=25\,934, \qquad N_{\rm gal}^{\rm total}=104\,858,\] so there are $N_{\rm gal}^{\rm non}=78\,924$ non-void galaxies in the same volume (Table~\ref{tab:number}; Figs.~\ref{fig:1}–\ref{fig:2}). 
The observed void galaxy fraction, $N_{\rm gal}^{\rm void} / N_{\rm gal}^{\rm total} \simeq 24.7\%$ is similar to the 20\% mentioned in other literature \citep{2019MNRAS.482.4329P}. We note that using a redshift-limited denominator ($z < 0.107$) appropriate for the \citet{2012MNRAS.421..926P} analysis volume yields a void fraction of $\sim 19.7\%$, consistent with our inference and reconciling the difference from the broadly quoted $11\%$.

Because SDSS $3''$ fibres sample only the inner regions of distant galaxies,
stellar masses are supplemented by the MPA–JHU catalogue 
\citep{2003MNRAS.341...33K,2004ApJ...613..898T}
, which ensures consistency with our SDSS spectroscopic anchor sample. For void galaxies we obtain 

\begin{eqnarray}
 \langle M_{\star}\rangle_{\rm void} &=&
 8.20^{+0.27}_{-0.24}\times10^{9}\,M_\odot,\\
 \langle{\rm SFR}\rangle_{\rm void} &=&
 0.71^{+1.52}_{-0.27}\,M_\odot\,{\rm yr^{-1}},
 \label{eq2}
 \\
 \langle{\rm sSFR}\rangle_{\rm void} &=&
 0.088^{+0.185}_{-0.033}\,{\rm Gyr^{-1}},
 \label{eq3}
\end{eqnarray}
where angle brackets denote sample means. % <— attention to usage after angle brackets
By removing the void-galaxy subset from the MPA–JHU cross-match, we obtain the corresponding non-void control sample:

\begin{eqnarray}
 \langle M_{\star}\rangle_{\rm non} &=&
 3.40^{+0.33}_{-0.30}\times10^{10}\,M_\odot,\\
 \langle\mathrm{SFR}\rangle_{\rm non} &=&
 1.34^{+1.60}_{-0.34}\,M_\odot\,\mathrm{yr^{-1}},
 \label{eq5}
 \\
 \langle\mathrm{sSFR}\rangle_{\rm non} &=&
 0.039^{+0.047}_{-0.011}\,\mathrm{Gyr^{-1}}.
 \label{eq6}
\end{eqnarray}

We confirm that, in this sample, void galaxies display the expected traits—lower stellar mass, lower metallicity, and higher specific SFR—compared with galaxies in denser regions \citep{2015ApJ...810..165L,2024A&A...692A.258A}. 

For comparison, \citet{2003MNRAS.341...33K} report a median stellar mass
$\langle M_{\star}\rangle_{\rm all}\simeq10^{10.5}\,M_\odot$ ($\approx3.16\times10^{10}\,M_\odot$)
for $122{,}808$ SDSS galaxies, consistent with (though slightly lower than) our non-void mean; the difference is expected because their value includes both void and non-void systems.

From these means we infer
\begin{equation}
 \langle M_{\star}\rangle_{\rm non}\;\simeq\;4\,\langle M_{\star}\rangle_{\rm void}.
\end{equation}
Combining this factor with the relative number counts gives the stellar–mass fraction in void galaxies,
\begin{equation}
 f_{\star}^{\rm void}
 \equiv 
 \frac{N_{\rm gal}^{\rm void}\,\langle M_{\star}\rangle_{\rm void}}
 {N_{\rm gal}^{\rm void}\,\langle M_{\star}\rangle_{\rm void}
 + N_{\rm gal}^{\rm non}\,\langle M_{\star}\rangle_{\rm non}}
 \;\approx\;0.10,
\end{equation}
i.e., void galaxies contribute $\sim10\%$ of the stellar–mass budget within this volume.

\subsection{Transient Samples}\label{subsec:transient_samples}

\subsubsection{Supernovae}
We include every supernova subtype in our analysis. Core-collapse supernovae—Types II, Ib, and Ic—trace massive, short-lived progenitors \citep{2003ApJ...591..288H, 2009ARA&A..47...63S}, whereas Type Ia events arise from delayed thermonuclear explosions of carbon–oxygen white dwarfs \citep{2000ARA&A..38..191H, 2014ARA&A..52..107M}.
Our supernova sample is drawn from the Sternberg Astronomical Institute (SAI) Supernova Catalogue \citep{1993BICDS..42...17T}, a comprehensive database that provides classifications, light curves, and host-galaxy information. Its chief advantage is rapid access to large numbers of supernovae with well-defined host data—crucial for comparing galaxy properties across contrasting cosmic environments.

Within the same three-dimensional volume defined for our void and non-void galaxy samples (Section 2.1), the catalogue lists 1783 supernovae ($\sum N_{\mathrm{Total,SN}} = 1783$; see Table \ref{tab:void_sn} and \ref{tab:non-void_sn}). These events form the foundation of our void–versus–non-void investigation. Our screening process revealed that a number of supernova events are located in host galaxies outside the coverage of the SDSS Main Galaxy Sample. Consequently, these supernovae have been excluded from the sample. Although some supernovae in the appendix table lack documented host galaxies, they still have measured redshifts. These redshifts are obtained from spectroscopic observations of the host galaxies. However, they can also be derived directly from the SN spectrum or estimated using photometric methods like light curve fitting.

\begin{table*}[htbp]
\centering
\caption{Counts of host galaxies and supernovae by environment within our VoidFinder volume ($0.005<z<0.05$; $90^\circ<\mathrm{RA}<270^\circ$, $-4^\circ<\mathrm{Dec}<71^\circ$). 
\emph{Galaxies} lists the number of SDSS spectroscopic galaxies in each environment. 
SN counts (from the SAI Supernova Catalogue, matched to these hosts) are split into thermonuclear Type~Ia and core–collapse (CCSNe; Types II/Ib/Ic); \emph{Unclassified} comprises events lacking a secure subtype. 
\emph{Total} is the sum of all SNe per environment. 
Quoted fractions in §\ref{section 3} are derived from these totals; 1$\sigma$ count uncertainties are Poisson ($\sqrt{N}$).}
\label{tab:number}
\begin{tabular}{lcccccc}
\toprule
 & Galaxies & Type-Ia SNe & CCSNe & Unclassified & Total & CCSNe fraction \\
\midrule
Inside Voids & 25\,934 & 144 & 246 & 40 & 430 & 57.21\% \\
Outside Voids & 78\,924 & 637 & 542 & 174& 1\,353 & 40.06\% \\
\bottomrule
\end{tabular}
\end{table*}

\begin{figure*}
 \vspace{0cm}
 \centering
 \subfigbottomskip=5pt
 \subfigcapskip=2pt
 \subfigure{\includegraphics[angle=0,scale=0.6]{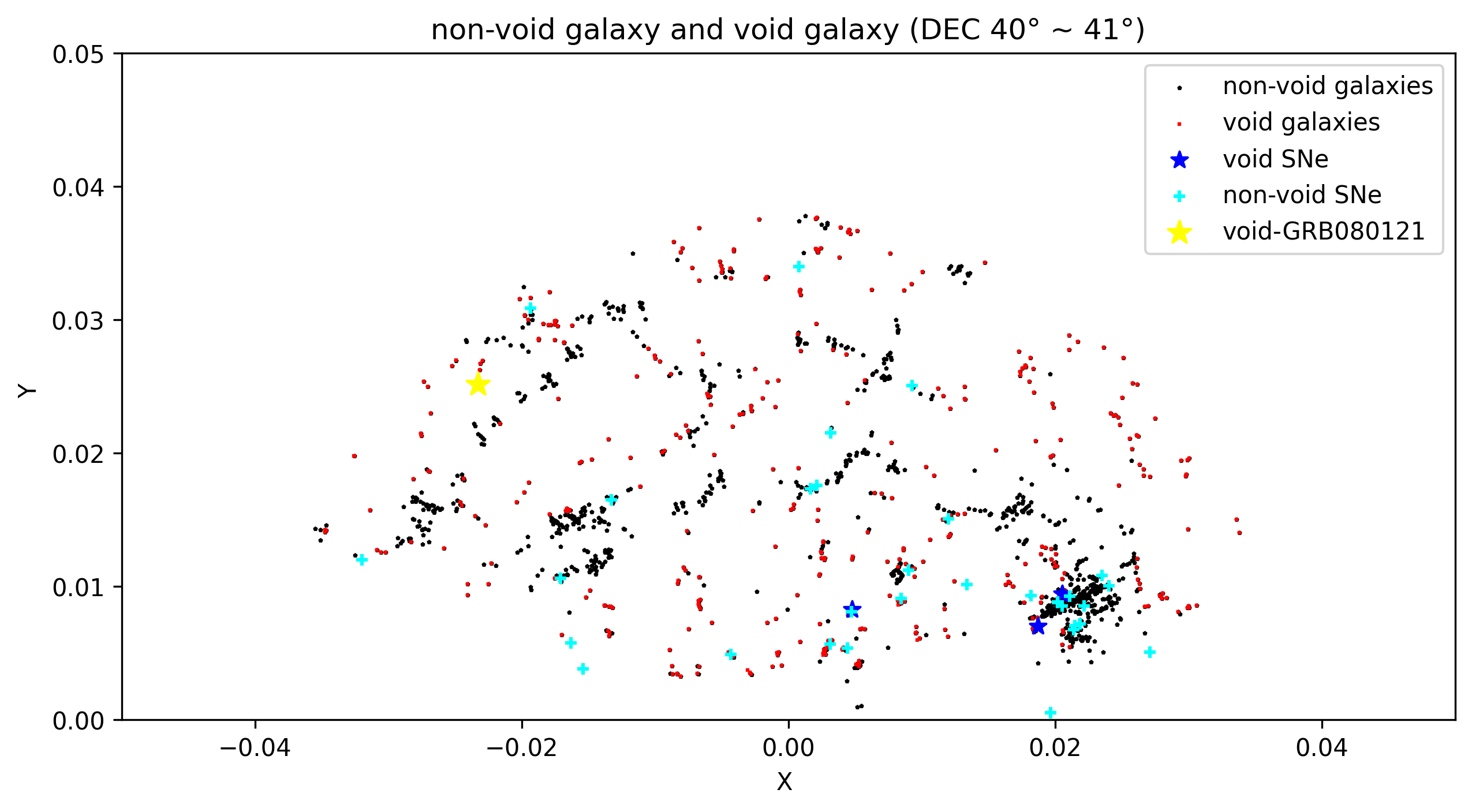}}
 \caption{Projected comoving maps for the declination slice within our VoidFinder volume: 
 $40^\circ$–$41^\circ$ which is chosen to include the only one GRB event in our sample.
Black points denote non-void galaxies; red points denote void galaxies; blue stars mark SNe in voids; cyan plus signs mark SNe in non-void regions. Yellow stars indicate GRBs: GRB\,080121. Axes $(X,Y)$ are projected comoving coordinates in the plane of each slice (see §\ref{section 2}); the numerical scale is set by the redshift distribution within each declination bin.}

\label{fig:1}
\end{figure*}

\begin{figure*}
 \vspace{0cm}
 \centering
 \subfigbottomskip=5pt
 \subfigcapskip=2pt
 \subfigure{\includegraphics[angle=0,scale=0.6]{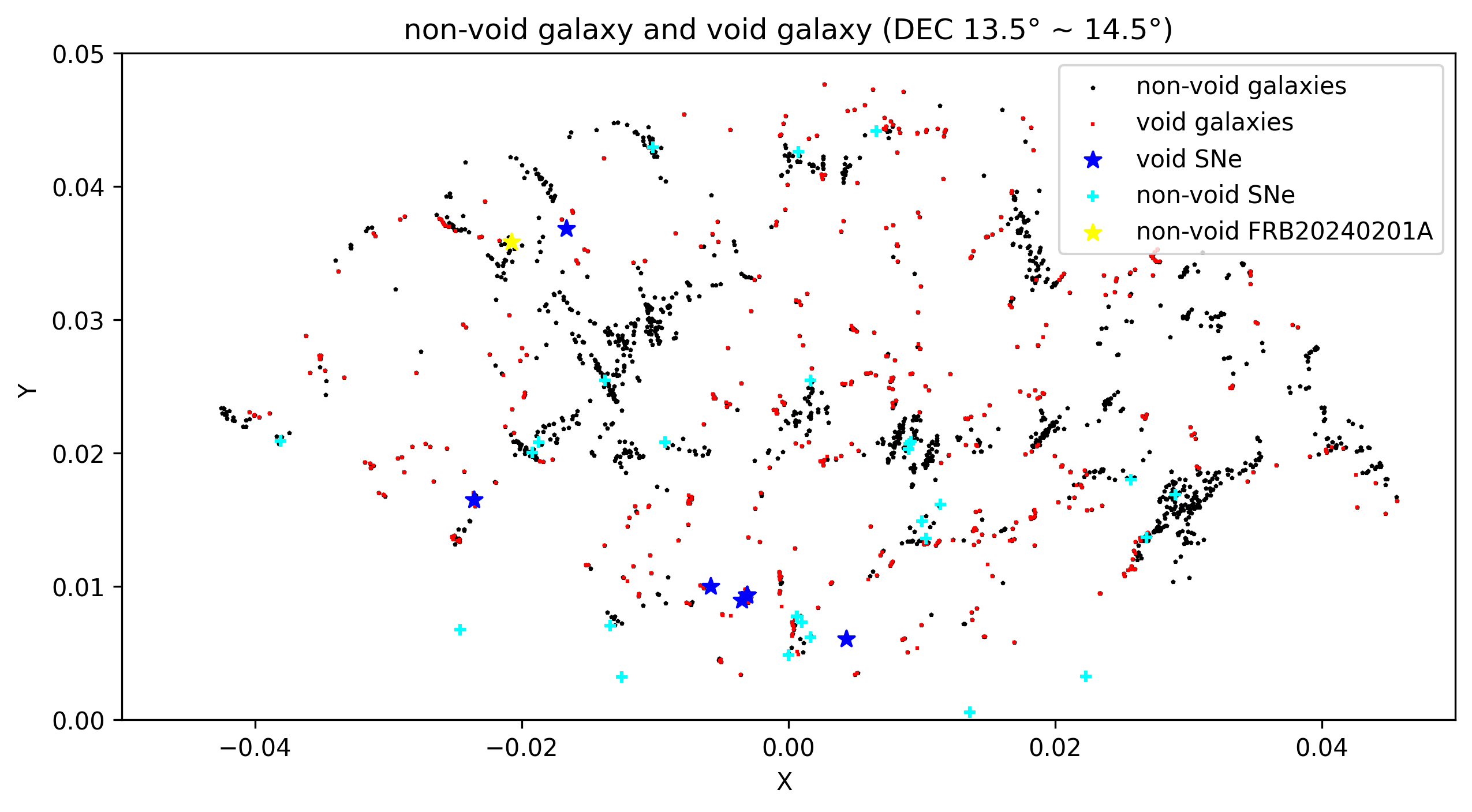}}
 \quad 
 \subfigure{\includegraphics[angle=0,scale=0.6]{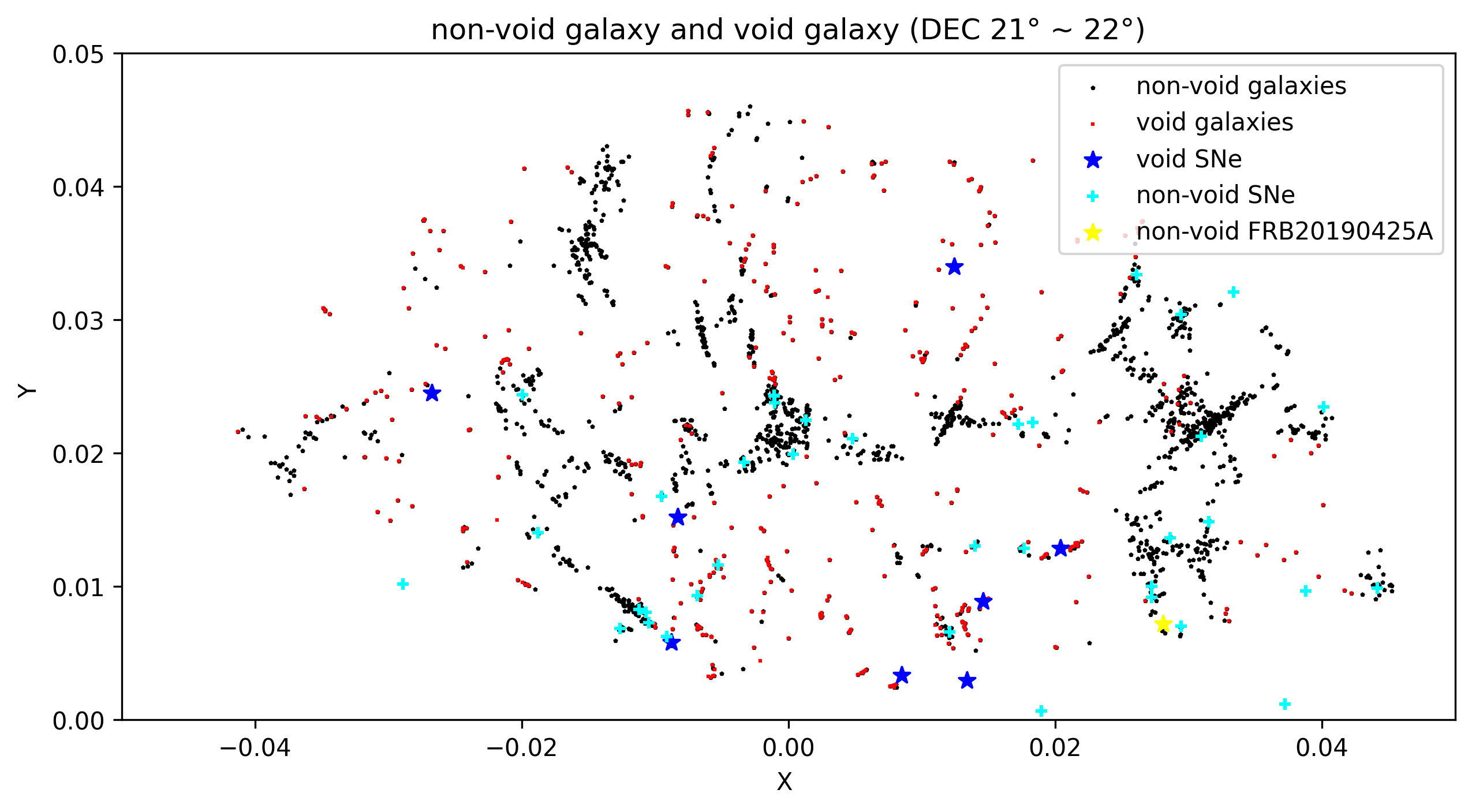}}
 \quad 
 \subfigure{\includegraphics[angle=0,scale=0.6]{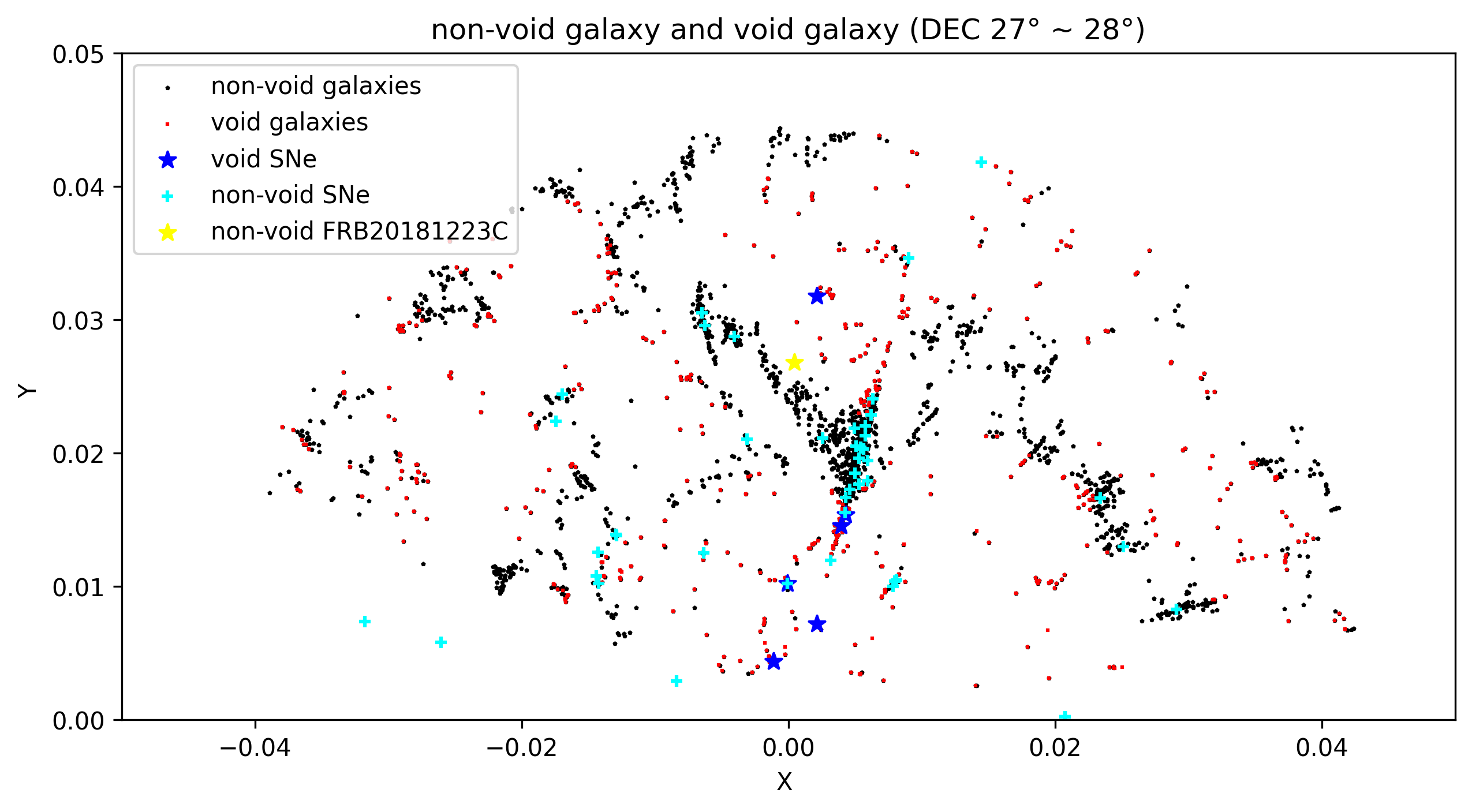}}

\caption{Projected comoving maps in three declination slices showing the distributions of void/non-void galaxies and transients. 
Panels (top to bottom) correspond to ${\rm DEC}=13.5^\circ$–$14.5^\circ$, $21^\circ$–$22^\circ$, and $27^\circ$–$28^\circ$ which are chosen to include the three FRB events in our sample. 
Symbols: black points—non-void galaxies; red points—void galaxies; blue stars—SNe in voids; cyan plus signs—SNe in non-void regions; yellow stars—FRBs located \emph{outside} voids (FRB\,20240201A, FRB\,20190425A, and FRB\,20181223C). 
Axes $(X,Y)$ are projected comoving coordinates in the plane of each slice (see §\ref{section 2}). 
No FRBs are identified within voids in our volume.}
\label{fig:2}
\end{figure*}

\subsubsection{Gamma-Ray Bursts}\label{subsec:grb_sample}
%-------------------------------------------------
We began by collecting \emph{all} gamma-ray bursts (GRBs) with secure redshift
measurements. 
Long-duration (Type II) GRBs originate from the collapse of rapidly rotating,
often stripped, massive stars
\citep{1993ApJ...405..273W,1999ApJ...524..262M,2006ARA&A..44..507W}, 
whereas short-duration (Type I) GRBs are produced by compact-binary mergers
(NS–NS or NS–BH)
\citep{1986ApJ...308L..43P,1989Natur.340..126E,1991AcA....41..257P}. 
These two classes therefore probe distinctly different stellar environments.

All sky coordinates and redshifts were taken from the on-line GRB catalogue
maintained by Jochen~Greiner at the Max Planck Institute for Extraterrestrial
Physics\footnote{\url{https://www.mpe.mpg.de/~jcg/grbgen.html}}. 
The database collates bursts discovered by \textit{BeppoSAX}, BATSE/RXTE,
ASM/RXTE, the IPN, HETE-II, \textit{INTEGRAL}, \textit{Swift}, AGILE,
\textit{Fermi}, MAXI, LEIA, \textit{Einstein Probe}, SVOM, SRG, and other
missions, and records both X-ray/optical afterglows and spectroscopic
redshifts. It thus provides a uniform, up-to-date resource for our
void–versus–non-void comparison. Redshift measurements for GRBs are generally acquired via spectroscopy of either their host galaxies or their subsequent optical afterglows.

Applying our redshift window, \(0.005 < z < 0.05\), yields seven GRBs.
Restricting the sample further to the right-ascension and declination limits
defined for the galaxy catalogue reduces the usable set to one event
(Table~\ref{tab:grb_info}):
\begin{itemize}
 \item \textbf{GRB~080121} (\(z=0.046\); short/Type I): With only a single GRB in our sample, this observation serves as a consistency check that compact-object mergers also occur in under-dense (void) galaxies—systems that are otherwise typical and expected to host NS–NS/NS–BH progenitors. Given the very large number of void galaxies relative to our single detection, the implied event rate—albeit with large uncertainties—is fully consistent with standard stellar-evolution expectations (see Table~\ref{tab:densities_rates_all}).
\end{itemize}

As observed by \textit{Swift}/BAT, the burst was a single peak with $T_{90} = 0.7 \pm 0.2$ s. Because of the weakness of the burst, a power-law fit to the spectrum from $T-0.4$ to $T+0.4$ seconds is not well constrained. The power-law index is $2.6 \pm 0.8$. The fluence was $(3 \pm 2) \times 10^{-8}~\mathrm{erg\,cm^{-2}\,s^{-1}}$. \citep{2008GCN..7209....1C}. Its prompt gamma-ray phase consists of a single pulse of $\sim0.2 $ s with a faint precursor emission visible in the soft energy range ($<$50 keV) \citep{2020MNRAS.492.5011D}. XRT and UVOT observations began at $\sim 2.3$~days after the GRB, and no credible counterpart was detected \citep{2008GCNR..118....1T}. At a redshift of $z = 0.046$, the isotropic energy release would be $\sim 10^{47}\ \mathrm{erg}$ \citep{2008GCN..7210....1P}.
Its host galaxy is an early-type galaxy with $M_B = -18.8~\rm mag$. From the MPA–JHU cross-match, the mass, SFR, and sSFR of the host are $M_\star = 2.29 \times 10^9~M_\odot, \quad \mathrm{SFR}_\star = 0.86~M_\odot~\mathrm{yr}^{-1}, \quad \mathrm{sSFR}_\star = 0.375~\mathrm{Gyr}^{-1}$, respectively. The burst has an angular separation of 28.64 arcsec from the center of its host galaxy, which corresponds to a projected physical offset of 25.6 kpc.

Although the sample is small, the presence of one burst in the under-dense environment allows a first, if tentative, assessment of the occurrence of GRB in voids versus walls.

%9.36	-0.06447061

\subsubsection{Fast Radio Bursts}

Fast radio bursts (FRBs) are bright, millisecond-duration radio transients most likely associated with magnetar activity, although other progenitor channels have also been proposed \citep{2007Sci...318..777L, 2020Natur.587...59B, 2013PASJ...65L..12T, 2016Natur.531..202S}. The majority of burst positions in our analysis are sourced from the CHIME/FRB Public Catalog\footnote{\url{https://www.chime-frb.ca/catalog}} \citep{2021ApJS..257...59C}, which provides sky coordinates, dispersion measures (DMs), and fluences. To select FRBs with secure \textbf{host-galaxy} identifications and redshift measurements, we cross-match these entries with the compilation of \citet{2025JCAP...01..036A}, who gathered all spectroscopically confirmed or robustly inferred FRB redshifts published to date and analyzed the redshift–DM relation; in our workflow, this compilation is the source of host identifications and redshifts, and redshifts in our sample are host-based, with DM used only as a descriptive observable. Together, these resources yield positions, DMs, and redshifts for 65 FRBs, providing the foundation for our statistical comparison of their spatial distribution and energetics within and outside the voids.

Within our adopted redshift range ($0.005 < z < 0.05$), we initially identify 16 FRBs. After applying the same right ascension and declination constraints used for the supernova and GRB samples, only three FRBs remain in the final sample. The positions of these three FRBs are shown in Figure~\ref{fig:2}, and the information of host galaxies are listed in Table~\ref{tab:frb_info} \citep{2025JCAP...01..036A,2024ApJ...971L..51B,2025PASA...42...36S}.

\subsection{Environmental Classification of Transients}
\label{subsec:env_classification}

Our analysis utilizes a pre-compiled catalog of void galaxies—systems definitively located within the underdense regions of the cosmic web. For each transient, we retrieved the host galaxy’s SDSS \texttt{objID} and cross-matched it against the VoidFinder void–galaxy catalog. A transient is labeled \emph{in-void} when its host appears in both databases; in that case we record the host name, the corresponding VoidFinder galaxy ID, and the SDSS \texttt{objID}. If no match is found, the transient is labeled \emph{outside-void}, and we retain only the host name and SDSS \texttt{objID}.

Quantifying the number of supernovae located inside and outside the voids, we find that $\mathbf{\sum \rm N_{\rm in,SNe} = 430}$ SNe reside within void regions, while $\mathbf{\sum \rm N_{\rm out,SNe} = 1353}$ occur in non-void environments. These SNe are further categorized by their types into Type Ia supernovae, core-collapse supernovae (CCSNe), and unclassified events, with a complete breakdown provided in Table~\ref{tab:number}. To examine their spatial distribution relative to void boundaries, we cross-referenced the SN positions with the distribution of void galaxies, as illustrated in Figures~\ref{fig:1} and~\ref{fig:2}.

Extending this analysis to gamma-ray bursts, we identify one notable case, GRB\,080121, located within a void region. The declination ranges of both void-associated guided the selection criteria depicted in Figure~\ref{fig:1}, with the complete GRB sample details summarized in Table~\ref{tab:grb_info}.

When applying the same methodology to fast radio bursts (FRBs), we find no events associated with void environments. Their spatial distribution (Figure~\ref{fig:2}), analyzed using the same framework as for SNe and GRBs, further confirms the absence of detected FRBs in cosmic voids. Taken at face value, this cross-class comparison shows an apparent contrast—SNe and GRBs exhibit measurable void associations in our sample, whereas no FRBs are presently identified in voids—but we treat this as a provisional, sample-limited observation rather than evidence for an intrinsic environmental preference. This apparent contrast may reflect the limited FRB sample size and the completeness of host identifications, which constrain our ability to characterize FRB distributions in voids. In addition, the sky coverage of the SDSS (DR7/DR8) footprint restricts our assessment of void associations to a non-universal volume.

\subsection{Density and Event Rate Calculation}
The relative density of transients with respect to galaxies is calculated by dividing the number of observed transient events by the total number of galaxies:
\begin{equation}
 n=\frac{\sum N_{\mathrm{Transient}}}{\sum N_{\mathrm{Galaxy}}}.
\end{equation}

The mass-normalized density (mass density) of transients is then determined by dividing the number density by the average stellar mass per galaxy:
\begin{equation}
 \rho=\frac{n}{m_{\mathrm{ave}}}.
\end{equation}

Finally, the volumetric event rate ($\Gamma$) is defined as:
\begin{equation}
 \Gamma = \frac{N}{\epsilon \cdot V \cdot T},
 \label{eq:event_rate_corrected}
\end{equation}
where $N$ represents the number of observed transient events, $V$ is the effective survey volume, $T$ denotes the total observation period, and $\epsilon$ is the observational duty cycle (the fraction of actual effective observation time relative to the total available time, with $0 \leq \epsilon \leq 1$). \textcolor{black}{We estimate the statistical uncertainties of the observed transient counts using Poisson statistics. For low-count statistics, we adopt the asymmetric 1 $\sigma$ Poisson confidence intervals following \citealp{1986ApJ...303..336G}.}

\begin{table*}[htbp]
\centering
\caption{Gamma–ray burst (GRB) sample within our survey volume. 
Columns list J2000 equatorial coordinates (RA, Dec), redshift, discovery/instrument, phenomenological class (Type~I = short/merger; Type~II =long/collapsar), host-galaxy identifier and morphology, projected offset from the host center (kpc), and the large–scale environment (void or non-void) assigned using the criterion in \S\ref{subsec:env_classification}. Only \textcolor{black}{one} GRB satisfy our cuts ($0.005<z<0.05$ and the RA/Dec footprint): GRB~080121 (Type~I; void).}
\label{tab:grb_info}
\begin{tabular}{lccccccccc}
\toprule
Name & RA ($^\circ$) & DEC ($^\circ$) & z & Source & Type & void ID/SDSS objID & Morphology & offset & In/Out? \\ 
\midrule
GRB 080121 & 137.23 & 41.84 & 0.046 & Swift/BAT & I & 18439/1237657630051467464 & \textcolor{black}{E/S0} & 25.6 & In \\
\bottomrule
\end{tabular}
\end{table*}

\begin{table*}[htbp]
\centering
\caption{Fast radio burst (FRB) sample within our survey footprint and redshift window ($0.005<z<0.05$). Columns list J2000 coordinates (RA, Dec), redshift, identified host galaxy and morphology (when available), and the large–scale environment (void or non-void) assigned following §\ref{subsec:env_classification}. 
All three FRBs in this volume lie in non-void regions; no FRBs are identified within voids. 
Positions/redshifts are compiled primarily from the CHIME/FRB Public Catalog, with host/redshift updates where available from \citet{2025JCAP...01..036A,2024ApJ...971L..51B,2025PASA...42...36S}.}
\label{tab:frb_info}
\begin{tabular}{lccccccccc}
\toprule
Name & RA ($^\circ$) & DEC ($^\circ$) & z & Host/SDSS objID & Morphology & In/Out?\\
\midrule
FRB 20181223C & 180.92 & 27.55 & 0.03024 & SDSS~J120340.98+273251.4/ 1237667323255455810 & \textcolor{black}{S} & Out \\ 
FRB 20190425A & 255.66 & 21.58 & 0.03122 & UGC10667/ 1237662474237575740 & \textcolor{black}{S (maybe Scd)} & Out \\ 
FRB 20240201A & 149.91 & 14.09 & 0.0427 & WISEA~J095937.44+140519.4/ 1237671123767066952 & \textcolor{black}{S} & Out \\ 
\bottomrule
\end{tabular}
\end{table*}

\begin{figure}
 \includegraphics [angle=0,scale=0.4] {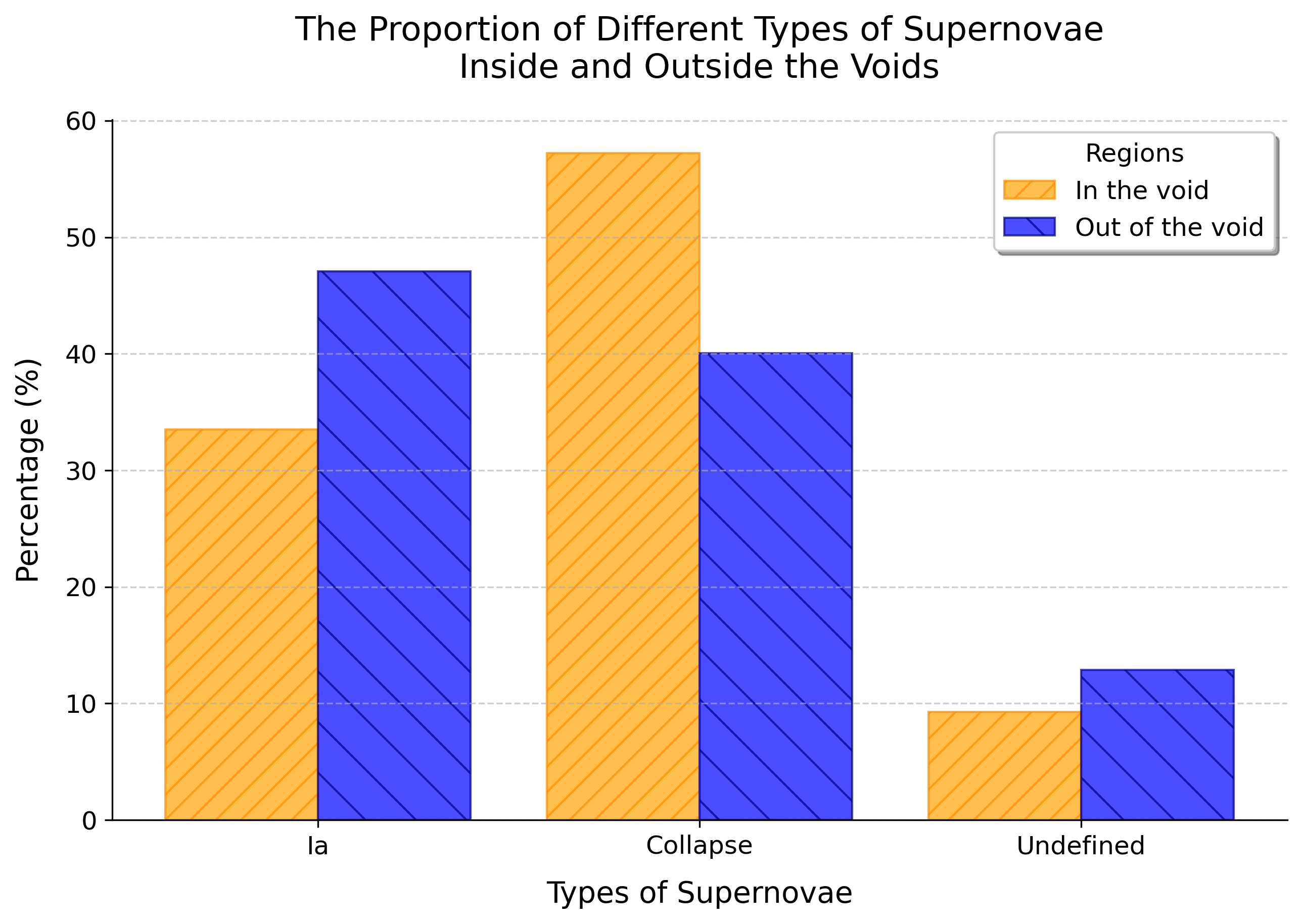}
 \centering
 \caption{Fractions of supernova subtypes inside and outside cosmic voids.
Yellow hatched bars mark SNe in voids; blue bars mark SNe in non-void regions. 
Categories are Type~Ia, core–collapse (labeled “Collapse”; Types II/Ib/Ic), and Unclassified. 
Values are the percentage of all SNe in each environment (void: Ia 33.49\%, CCSNe 57.21\%, Unclassified 9.30\%; non-void: Ia 47.08\%, CCSNe 40.06\%, Unclassified 12.86\%), computed from Table~\ref{tab:number}.}

 \label{fig:3}
\end{figure}

%-----

\section{Statistical Results}
\label{section 3}

Based on above samples, we systematically analyze the properties of SNe, GRBs, and FRBs within and outside cosmic voids, comparing their number density, mass density, and event rates to uncover fundamental differences between void and non-void environments.

\subsection{Supernova Fractions}

Using data from Table~\ref{tab:number}, we find that the number of Type Ia SNe inside and outside voids is $\sum \rm N_{in,Ia}=144$ and $\sum \rm N_{out,Ia}=637$, respectively. The fraction of Type Ia SNe within voids is
\begin{equation}
\frac{\sum \rm N_{in,Ia}}{\sum \rm N_{in,SNe}}=0.3349,
\end{equation}
while the corresponding fraction outside voids is
\begin{equation}
\frac{\sum\rm N_{out,Ia}}{\sum\rm N_{out,SNe}}=0.4708.
\end{equation}

Using the same method, we calculate the fractions for core-collapse supernovae (CCSNe) as follows:

\begin{equation}
\frac{\sum \rm N_{in,CCSNE}}{\sum \rm N_{in,SNe}}=0.5721,
\end{equation}
\begin{equation}
\frac{\sum\rm N_{out,CCSNE}}{\sum\rm N_{out,SNe}}=0.4006.
\end{equation}

These results are visualized in Figure~\ref{fig:3}, where a notable pattern emerges: the fraction of Type~Ia SNe is lower in voids than in non-void environments, whereas CCSNe are more prevalent in void regions. To investigate the origin of this trend, we further analyzed the morphological types of SN host galaxies. As shown in Figure~\ref{fig:4}, the distribution indicates a shift from early-type to late-type galaxies, which typically host younger stellar populations. This suggests that the higher CCSN \emph{fraction} in voids may be linked to a greater proportion of star-forming galaxies in these underdense regions, rather than to changes in the explosion channels themselves. A more detailed interpretation of this correlation is provided in Section~\ref{section 4}, where we explore its implications for star formation and galaxy evolution in low-density environments.

\subsection{Number and Mass Density}

In addition to fractions, we estimated the number density, mass density, and event rates of SNe within and outside cosmic voids. The number density of supernovae relative to galaxies is calculated as:
\begin{equation}
n_{\rm in}=\frac{\sum\rm N_{in,SNe}}{\sum\rm N_{in,galaxy}}=0.01658,
\end{equation}
and
\begin{equation}
n_{\rm out}=\frac{\sum\rm N_{out,SNe}}{\sum\rm N_{out,galaxy}}=0.01714.
\end{equation}
The mass density was obtained by combining the number density with the average stellar mass per galaxy (see Eq. 2), yielding:

\begin{equation}
 \rho_{\rm in}=\frac{n_{\rm in}}{m_{\rm in,ave}}=2.02^{+0.07}_{-0.06} \times 10^{-12} \rm M_{\odot}^{-1}
\end{equation}
and\textcolor{black}{
\begin{equation}
 \rho_{\rm out}=\frac{n_{\rm out}}{m_{\rm out,ave}}=5.04^{+1.47}_{-1.45} \times 10^{-13} \rm M_{\odot}^{-1}
\end{equation}}

Although the galaxy count within voids is relatively low, their supernova production proportionally exceeds that of galaxies outside voids. The supernova rate per unit mass appears to be higher inside void regions compared to outside.

Table~\ref{tab:densities_rates_all} further presents the number and mass densities for Type Ia and CCSNe. CCSNe densities inside voids are roughly double those of Type Ia SNe. Ia SNe have lower number densities in voids than in denser environments, but CCSNe is the opposite. However, the mass density inside void of Type Ia SNe is more than twice that outside void, while the mass density inside void of CCSNe is more than five times that outside void.

The spatial distribution of SNe exhibits clear environmental dependence, reflecting their sensitivity to large-scale structure. Type Ia SNe are especially influenced by local host environmental conditions, which affect progenitor accumulation timescales. Intriguingly, the anomalously high mass density of CCSNe in voids suggests that their massive progenitors ($>8\,\mathrm{M_\odot}$) may preferentially form in low-density environments. This could result from (i) higher formation efficiency—more CCSNe per unit stellar mass—possibly reflecting an initial mass function (IMF) biased toward higher stellar masses in low-metallicity conditions; however, our data do not directly constrain the IMF, (ii) a younger stellar population (recent concentrated star formation), or (iii) unique star-forming conditions in voids, such as low metallicity, which is known to favor the formation of massive stars. 

Based on a standard Kroupa/Chabrier IMF, a simple order-of-magnitude estimate yields: the core-collapse supernova rate scales with star formation rate as $R_{\rm CCSN} \approx \eta_{\rm SN} \, \mathrm{SFR}$, and the rate per unit stellar mass as $r_{\rm CCSN} \approx \eta_{\rm SN} \, \mathrm{sSFR}$, where $\eta_{\rm SN}$ is the IMF-dependent yield of $>8\,M_\odot$ progenitors per unit formed stellar mass. Since $\eta_{\rm SN}$ remains constant under a standard IMF, the observed enhancement in sSFR in void galaxies directly predicts a commensurate increase in CCSN incidence, consistent with Table~\ref{tab:densities_rates_all} within uncertainties\footnote{Our order-of-magnitude analysis shows that the elevated sSFR in void galaxies is sufficient to explain the higher CCSN rate, making an altered IMF an unnecessary, more speculative explanation. The data in Table~\ref{tab:densities_rates_all} are consistent with this straightforward interpretation.}.

As a preliminary check, we compare CCSN occurrence in bins of host stellar mass and sSFR (Sec.~3.1) and find qualitatively consistent trends within uncertainties. While some aspects of these explanations can be tested with the available host-galaxy properties (e.g., sSFR, stellar mass, metallicity), a full investigation of the relative contributions of these factors is left for future work.

Although our GRB and FRB samples are limited—comprising only one GRB and no FRBs associated with void regions—we still report their number densities and mass-normalized occurrence rates, as listed in Table \ref{tab:densities_rates_all}. However, these small-number statistics preclude meaningful comparison with the supernova population and limit our ability to draw robust conclusions about their environmental dependence based on the current dataset.

\begin{table*}[htbp]
\centering
\caption{Number / Mass-Normalized Densities and Event Rates of Transients in Void vs.\ Non-void Environments}
\label{tab:densities_rates_all}
\begin{tabular}{l l c c c}
\toprule
\multirow{2}{*}{Class} & \multirow{2}{*}{Quantity} & \multicolumn{2}{c}{Environment} & \multirow{2}{*}{Benchmark / Reference}\\
\cmidrule(lr){3-4}
 & & Inside Voids & Outside Voids & \\
\midrule
\multirow{4}{*}{Type Ia SNe}
 & Count & 144 & 637 & \multirow{3}{*}{—} \\
 & Number density ($\mathrm{gal^{-1}}$) & 0.00555 & 0.00807 & \multirow{3}{*}{—} \\
 & Mass-normalized density ($M_\odot^{-1}$) & ${6.77^{+0.22}_{-0.20}\!\times\!10^{-13}}$ & ${2.37^{+0.23}_{-0.21}\times10^{-13}}$ & \\
 & Event rate ($\mathrm{Gpc^{-3}\,yr^{-1}}$) & $2.31^{+0.29}_{-0.29} \times10^{3}$ & ${2.50^{+0.31}_{-0.31} \times10^{4}}$ &
 $2.28^{+0.20}_{-0.20} \times10^{4}$ \citep{2024MNRAS.530.5016D} \\
\addlinespace[4pt]
\multirow{4}{*}{CCSNe}
& Count & 246 & 542 & \multirow{3}{*}{—} \\
 & Number density ($\mathrm{gal^{-1}}$) & 0.00949 & 0.00687& \multirow{3}{*}{—} \\
 & Mass-normalized density ($M_\odot^{-1}$) & ${1.16^{+0.04}_{-0.03}\!\times\!10^{-12}}$ & {${2.02^{+0.20}_{-0.18}\times10^{-13}}$} & \\
 & Event rate ($\mathrm{Gpc^{-3}\,yr^{-1}}$) & $3.90^{+0.52}_{-0.52}\times10^{3}$ & ${2.90^{+0.33}_{-0.33}\times10^{4}}$ &
 $7.0^{+1.0}_{-0.9}\times10^{4}$\citep{2025arXiv250810985P} \\
\addlinespace[4pt]
\multirow{4}{*}{GRBs\,$^{\dagger}$}
& Count & 1 & 0 & \multirow{3}{*}{—} \\
 & Number density ($\mathrm{gal^{-1}}$) & ${3.856\times10^{-5}}$ & $0$ & \multirow{1}{*}{—} \\
 & Mass-normalized density ($M_\odot^{-1}$) & $ {4.70^{+0.15}_{-0.14} \times10^{-15}}$ & {$0$} & \multirow{1}{*}{—} \\
 & Event rate: Type I / II ($\mathrm{Gpc^{-3}\,yr^{-1}}$) & $9.37^{+12.37}_{-1.22} \ /\ 0$ & $0\ /\ \textcolor{black}{0}$ & \shortstack{$101_{-80}^{+220}$ (sGRBs) \citep{2018NatCo...9..447Z},\\$164^{+68}_{-65}$ (LL LGRBs) \citep{2015ApJ...812...33S}} \\
\addlinespace[4pt]
\multirow{4}{*}{FRBs}
& Count & 0 & 3 & \multirow{3}{*}{—} \\
 & Number density ($\mathrm{gal^{-1}}$) & $0$ & $ {3.801\times10^{-5}}$ & \\
 & Mass-normalized density ($M_\odot^{-1}$) & $0$ & {${1.12^{+0.11}_{-0.10}\times10^{-15}}$} & \\
 & Event rate ($\mathrm{Gpc^{-3}\,yr^{-1}}$) & $0$ & $140.75$ & $10^{3}$–$10^{4}$ (DM-based)\citep{2020MNRAS.494..665L} \\
\bottomrule
\end{tabular}

\vspace{6pt}
\footnotesize\textit{Notes.} 
“Number density” is $\sum N_{\rm Transient}/\sum N_{\rm Galaxy}$.
“Mass-normalized density” divides the number density by the average stellar mass per galaxy in the corresponding environment (see §2). 
Event rates follow Eq.~(\ref{eq:event_rate_corrected}). 
\textcolor{black}{We adopt the date range of 2010–2012 in order to match that used by \citet{2024MNRAS.530.5016D,2025arXiv250810985P}, thereby enabling a direct and consistent comparison of the supernova event rates. Notably, the detection rate of SNe during this interval appears relatively stable.} $^{\dagger}$GRB densities are for the combined GRB sample; event rates are split into Type~I (short) and Type~II (long). Benchmark values for GRBs (sGRB and low-luminosity LGRB rates) and for FRBs (DM-based cosmic rates) are shown for context. The rate for ``host-confirmed (SDSS footprint, z $<$ 0.05) short GRBs" is selection-limited and should be considered a lower bound relative to the population-averaged local rates.
\end{table*}

\subsection{Event Rate}

Building upon the above analysis, we estimate the event rates of different transient populations to assess their occurrence frequencies across distinct cosmic environments. The number of transients inside and outside voids has already been established. The total volume is computed using three-dimensional spatial coordinates, with voids assumed to occupy 70\% of the total cosmic volume and non-void regions the remaining 30\% \textcolor{black}{\citep{2012MNRAS.421..926P,2014MNRAS.442..462S,2014MNRAS.441.2923C,2018MNRAS.473.1195L,2019MNRAS.487.1607G}.} For supernovae (SNe), we adopt a survey time of $T = 2012$–$2010$=3 yr, and assume a duty cycle $\epsilon$ of 100\% for simplicity. The resulting SN event rates are presented in Table~\ref{tab:densities_rates_all}.

Interestingly, we identify one Type~I (short) GRB consistent with a compact-binary (NS–NS/NS–BH) merger in a void. While population studies and models suggest that such mergers occur more often in massive, evolved hosts characteristic of denser environments \citep{2013ApJ...769...56F,2019MNRAS.487.1675A}, they are also expected—albeit less frequently—in low-density regions because of broad delay-time distributions and natal-kick–induced offsets. This single detection is therefore consistent with those expectations, and, given the sample size, the inferred rate is subject to large uncertainties. To investigate this further, we assume an operational period beginning in 2008 for \textit{Fermi}/GBM (with a duty cycle of 50\% \citep{2016ApJS..223...28N,2018NatCo...9..447Z}). Event rate estimates for the one GRB are summarized in \label{tab:densities_rates_all}. Although the inferred void GRB volumetric rate of $9.37^{+12.37}_{-1.22}~{\rm Gpc^{-3}~yr^{-1}}$ is low in absolute terms, it remains consistent within uncertainties with prior estimates of the local sGRB rate, such as $101^{+220}_{-80}~{\rm Gpc^{-3}~yr^{-1}}$ reported by \citealt{2018NatCo...9..447Z}. Given our limited sample volume and single-event detection, our result does not contradict existing expectations \citep{2014ARA&A..52...43B,2015MNRAS.448.3026W}. But our GRB rate is \emph{not} directly comparable to literature-wide, selection-averaged GRB rates due to ``host-confirmed (SDSS footprint, z $<$ 0.05) short GRBs". Moreover, the prompt emission properties of the detected GRB show no significant deviation from the broader Type I populations, suggesting that void environments may not strongly alter their observable characteristics.

We also estimate the event rate of fast radio bursts (FRBs) occurring outside voids, as no FRBs have yet been detected within void regions. This analysis spans observations from 2017 to 2025 and assumes a detector duty cycle of 90\% \citep{2018ApJ...863...48C,2021ApJS..257...59C}. The resulting FRB rate in non-void regions is found to be one to two orders of magnitude lower than the values predicted by dispersion measure (DM)-based cosmological estimates \citep{2020MNRAS.494..665L}. This discrepancy likely reflects differences in methodology: while DM-based models integrate over the entire cosmic volume and account for contributions from the intergalactic medium, our approach is restricted to a small, volume-limited sample with confirmed host associations. In particular, the strict requirement for redshift and host identification greatly limits the sample size, leading to an inherently lower inferred volumetric rate. At present, the absence of FRBs in voids is consistent with small-number statistics and the limited footprint/$z<0.05$ window of this analysis. While magnetar formation is linked to recent star formation and massive progenitors, the elevated CCSN rates and higher sSFRs observed in void hosts indicate that such progenitor channels may still operate there. A larger, uniformly localized FRB sample will be required to assess any genuine environmental trend.

\begin{figure*}
 \vspace{0cm}
 \centering
 \subfigbottomskip=20pt
 \subfigcapskip=2pt
 \subfigure{\includegraphics[angle=0,scale=0.42]{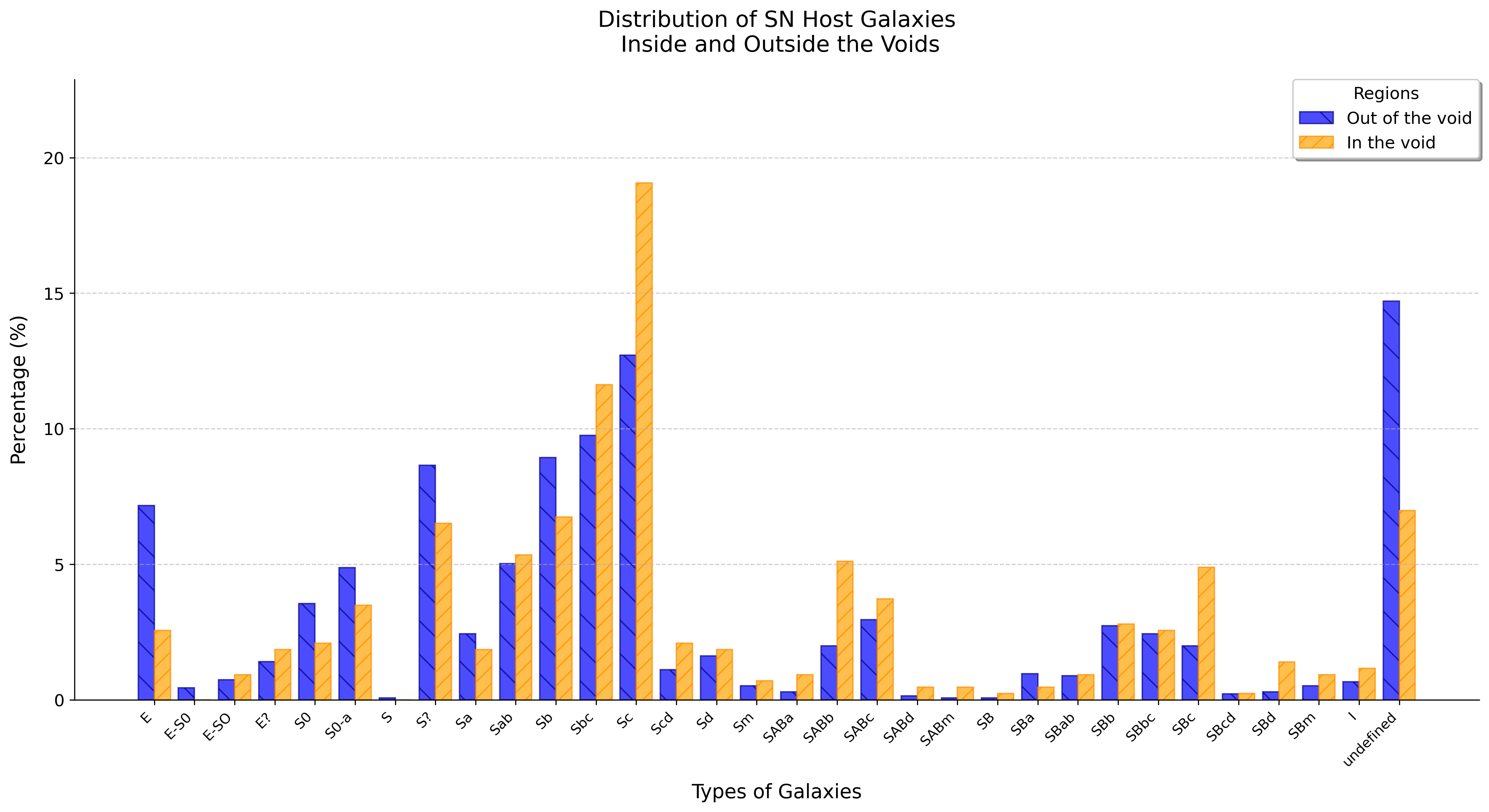}}
\caption{
Distribution of morphological types for supernova (SN) host galaxies located inside (orange) and outside (blue) cosmic voids. The x-axis follows the standard galaxy classification sequence: elliptical (E), lenticular (S0), spiral (Sa–Sd), barred spiral (SBa–SBm), irregular (I), and undefined. Spiral subclasses (a–m) indicate the degree of spiral arm winding, from tightly wound (`a') to loosely wound or irregular (`m'). Void SN hosts are skewed toward late-type spirals (e.g., Sc, Scd), while non-void SN hosts show a broader morphological mix, including more early-type systems.}
\label{fig:4}
\end{figure*}

\section{Conclusion and Discussion}
\label{section 4}

This work presents a systematic census of supernovae, gamma-ray bursts, and fast-radio bursts within well-defined nearby cosmic voids. By matching each transient to the large-scale structure traced by SDSS galaxies, we compare their volumetric rates, host demographics, and environmental preferences between underdense and average-density regions of the local Universe.

We find that core-collapse supernovae (CCSNe) dominate the transient inventory of voids, comprising about $57\%$ of all supernovae compared to $40\%$ outside voids, with a mass-normalized rate roughly five times higher. In contrast, Type~Ia supernovae are relatively suppressed, representing only $\sim33\%$ of void events but nearly half of those in denser regions. Void supernovae preferentially occur in late-type, star-forming hosts (Fig.~\ref{fig:4}), while early-type hosts dominate in non-void environments. In addition, one short GRB, GRB\,080121 (Type~I), is identified within a void and is hosted by a low-mass galaxy at $z=0.046$, consistent in prompt properties with the general short GRB population \citep{2022Univ....8..310P}. No FRBs are found within voids despite comparable sky coverage and redshift range; the non-void rate measured here remains one–to–two orders of magnitude below DM-based cosmic averages.

The enhanced CCSN rate in void galaxies can be attributed to their elevated specific star-formation rates (sSFRs) and low metallicities, which promote the formation of massive, short-lived progenitors. Although our data do not directly constrain the stellar initial mass function (IMF), it is plausible that metal-poor environments favor a top-heavy IMF, further amplifying CCSN production. The observed CCSN excess and Type~Ia deficit are therefore best explained by intrinsic host properties—lower metallicity and higher sSFR—rather than by any direct effect of large-scale structure on stellar evolution \citep{2015ApJ...810..165L,2024A&A...692A.258A}. Since SN~Ia progenitors require long delay times, their scarcity in voids likely reflects the lack of old, metal-rich stellar populations \citep{2023Natur.619..269D}. While large-scale density may indirectly influence star formation through gas accretion or interaction frequency, its direct role in shaping stellar lifecycles remains uncertain. In metal-poor, gas-rich disks, continuous replenishment of massive stars sustains a high CCSN yield, whereas the formation of SN~Ia progenitors proceeds less efficiently. If low metallicity also facilitates angular-momentum retention, future wide-field surveys may reveal a relative excess of broad-lined Type~Ic SNe and possibly long GRBs in nearby voids.

The host morphologies of void transients reinforce this interpretation. Late-type galaxies dominate in underdense regions, while early-type systems prevail in denser areas, reflecting the morphology–density relation \citep{1984ApJ...281...95P}. Morphology traces star formation and stellar mass: late-type galaxies exhibit younger populations, higher gas fractions, and elevated sSFRs, favoring CCSNe, whereas early-type galaxies are older, more massive, and quiescent, consistent with delayed SN~Ia progenitors. In our sample, the average SFR of non-void galaxies ($\sim1.34\ M_\odot\ {\rm yr^{-1}}$, Eq.~\ref{eq5}) exceeds that of void galaxies ($\sim0.71\ M_\odot\ {\rm yr^{-1}}$, Eq.~\ref{eq2}), yet their lower stellar masses yield higher sSFRs (Eqs.~\ref{eq3} and~\ref{eq6}). This explains the elevated CCSN frequency in voids and the relative suppression of SN~Ia events \citep{2024A&A...687A..98C}.

The presence of GRB\,080121 in a void implies that compact binaries may form in situ from local young stellar populations or survive natal kicks within low-mass gravitational potentials. However, this conclusion is limited by small-number statistics. Short GRBs typically exhibit projected offsets of $\sim5$~kpc from their hosts \citep{2010ApJ...708....9F,2016ApJS..227....7L}, but GRB\,080121 shows a larger offset ($\sim25.6$~kpc), possibly due to void expansion \citep{2022BAAA...63..193C,2025JCAP...05..011B} and the weak gravitational binding of low-mass hosts. Host metallicity may also influence long GRB progenitors \citep{2016ApJ...817....8P,2016A&A...590A.129J}; since galaxy mass, metallicity, and environment are closely linked \citep{2008MNRAS.390..245C,2015ApJ...805..121D}, the low-mass, metal-poor galaxies characteristic of voids may be promising sites for future long-GRB detections.

The absence of FRBs in voids likely arises from limited statistics and selection effects rather than an intrinsic deficit of magnetars. Localized FRB hosts span a wide range in stellar mass and SFR \citep{2020ApJ...895L..37B,2020ApJ...903..152H}, suggesting multiple progenitor channels. Our current sample is too small for statistical inference, but future wide-field radio surveys with improved localization will be essential to test whether FRBs truly avoid underdense regions.

Overall, our findings suggest that voids serve as dynamically young environments where galaxy evolution and transient production follow distinct pathways. The relative enhancement of CCSNe, suppression of SN~Ia events, and potential occurrence of GRBs all point to stellar populations that are younger, more metal-poor, and actively forming stars. As future transient surveys expand, the contrast between voids and denser regions will provide a powerful probe of how cosmic structure modulates the death of stars across the nearby universe.

\begin{acknowledgements}
We acknowledge the support by the National Key Research and Development Programs of China (2022YFF0711404, 2022SKA0130102), the National SKA Program of China (2022SKA0130100), the National Natural Science Foundation of China (grant Nos. 12573046, 12121003), the science research grants from the China Manned Space Project with NO. CMS-CSST-2021-B11, the Fundamental Research Funds for the Central Universities, and the Program for Innovative Talents and Entrepreneurs in Jiangsu. RGB acknowledges financial support from the Severo Ochoa grant CEX2021-001131-S funded by MCIN/AEI/ 10.13039/501100011033 and to grant PID2022-141755NB-I00.
\end{acknowledgements}

\appendix
This appendix presents two tables that list detailed information for supernovae located inside (Table \ref{tab:void_sn}) and outside (Table \ref{tab:non-void_sn}) cosmic voids, respectively. These data serve as the foundation for analyzing the distribution of supernovae and the properties of their host galaxies in different large-scale environments.

%*********************************************reference********************************************

%%%%%%%%%%%%%%%%%%%%%%%%%%%

\begin{center}
\makeatletter
\setlength{\LTcapwidth}{\textwidth}
\makeatother

\begin{longtable}{ccccccccccc}
\caption{\textcolor{black}{Supernovae sample inside the voids within our survey volume. Columns list SNe name, Host galaxy/void galaxy number \citep{2012MNRAS.421..926P}/SDSS objID, morphology of host galaxy.}\label{tab:void_sn}} \\
\hline
SN & Type & Host/void ID/SDSS objID & Morphology \\ \hline
\endfirsthead

\multicolumn{10}{c}{{\tablename\ \thetable{} -- continued from previous page}} \\
\hline
SN & Type & Host/void ID/SDSS objID & Morphology \\ \hline
\endhead

\hline
\multicolumn{10}{c}{{- To be continued}} \\
\endfoot

\hline
\endlastfoot

 1915A & ~ & NGC4527/68390/1237651754014081039 & SABb \\ 
 1920A & II: & NGC2608/68912/1237661126687588382 & Sb \\ 
 1926A & II & NGC4303/68470/1237668675085533189 & Sbc \\ 
 1938B & ~ & NGC2672/69500/1237667211044978721 & E \\ 
 1950I & ~ & MCG+03-39-008/63416/1237667967496159276 & S0-a \\ 
 1950M & ~ & NGC3266/68312/1237651272432615436 & S0 \\ 
 1952G & ~ & NGC5668/68532/1237654880738476143 & Scd \\ 
 1954B & I & NGC5668/68532/1237654880738476143 & Scd \\
\end{longtable}
\end{center}

%%%%%%%%%%%%%%%%%%%%%%%%%%%%%%%
%outside void SN

\begin{center}
\makeatletter
\setlength{\LTcapwidth}{\textwidth}
\makeatother

\begin{longtable}{ccccccccccc}
\caption{\textcolor{black}{Supernovae sample outside the voids within our survey volume. Columns list SNe name, Host galaxy/SDSS objID, morphology of host galaxy.}\label{tab:non-void_sn}} \\
\hline
SN & Type & Host/SDSS objID & Morphology \\ \hline
\endfirsthead

\multicolumn{10}{c}{{\tablename\ \thetable{} -- continued from previous page}} \\
\hline
SN & Type & Host/SDSS objID & Morphology \\ \hline
\endhead

\hline
\multicolumn{10}{c}{{- To be continued}} \\
\endfoot

\hline
\endlastfoot

 1901A & ~ & NGC2535/1237660764834234637 & Sc \\ 
 1901B & I & NGC4321/1237662526839914546 & SABb \\ 
 1914A & ~ & NGC4321/1237662526839914546 & SABb \\ 
 1926B & ~ & NGC6181/1237662698115367116 & SABc \\ 
 1936A & IIL & NGC4273/1237654880723927186 & Sc \\ 
 1940C & II & IC1099/1237654398091788391 & SABc \\ 
 1940D & ~ & NGC4545/1237655349952118906 & Sc \\ 
 1946A & ~ & NGC3977/1237658802573410379 & Sab \\
\end{longtable}
\end{center}
%%%%%%%%%%%%%%%%%%%%%%%%%%%%%%%

\end{document}